# Modeling System Events and Negative Events Using Thinging Machines Based on Lupascian Logic

Sabah Al-Fedaghi[*]

*Computer Engineering Department*
*Kuwait University*
*Kuwait*

salfedaghi@yahoo.com, sabah.alfedaghi@ku.edu.kw

*Abstract* - This paper is an exploration of the ontological foundations of conceptual modeling that addresses the concept of events and related notions. Development models that convey how things change over space and time demand continued attention in systems and software engineering. In this context, foundational matters in modeling systems include the definition of an event, the types of events, and the kinds of relationships that can be recognized among events. Although a broad spectrum of research of such issues exists in various fields of study, events have extensive applicability in computing (e.g., event-driven programming, architecture, data modeling, automation, and surveillance). While these computing notions are diverse, their event-based nature lets us apply many of the same software engineering techniques to all of them. In this paper, the focus is on addressing the dynamic concepts of system events and negative events. Specifically, we concentrate on what computer scientists would refer to as an *event grammar* and *event calculus*. Analyzing the concept of event would further the understanding of the event notion and provide a sound foundation for improving the theory and practice of conceptual modeling. An event in computer science has many definitions (e.g., anything that happens, changes in the properties of objects, and the occurrence of and transition between states). This paper is based upon a different conceptualization using thinging machines and Lupascian logic to define negative events. An *event* is defined as a time penetrated domain's region, which is described in terms of things and five-action machines. Accordingly, samples from event grammar and event calculus are remodeled and analyzed in terms of this definition. The results point to an enriched modeling technique with an enhanced conceptualization of events that can benefit behavior modeling in systems.

## I. INTRODUCTION

In an information society, flows of data can be seen as streams of observable events. The key idea is to explore temporal, causal, and semantic relationships among events to make sense of them in a timely fashion [1]. The interconnected nature of the modern world means incorporating developments related to event-processing technologies in various domains (e.g., smart healthcare).

---------------------------------------------------------

In this context, according to Hornsby et al. [2], the development of conceptual models that convey the ways objects change over space and time demands continued attention from software engineers. Events have broad applicability in computing, event-driven programming, and event-driven architecture. Although these computing notions are diverse, their event-based nature lets us apply many of the same software engineering techniques to all of them [3].

### A. Overview of Events in Fields of Study

A broad spectrum of research on the concept of event in various fields of study use many different definitions of event. In philosophy, events are viewed as the changes in the form of objects obtaining or losing their properties [4]. As a mainstream notion in contemporary philosophy, events remain a popular subject in various research areas [4]. Etzion and Niblett [5] defined an event as an occurrence within a particular system; it is something that has happened, or is contemplated as having happened in a domain. The word *event* is also used to mean a programming entity that represents such an occurrence in a computing system.

In software engineering, according to Auguston and Whitcomb [6], the concept of software models based on events and event traces was introduced as an approach to software debugging and testing automation. For example, event processing is an emergent area driven primarily by the greater need of enterprises to respond quickly to this large volume of business and IT events [7].

### B. Importance of Events for system Modeling

The definition of an event is one of the foundational matters in any system [5]. We are in urgent need of event-based theory, design methods, and tools [8]. Analyzing the concept of event would further the understanding of events and provide a sound foundation for the theory and practice of conceptual modeling. According to Guizzardi et al. [9], given the importance of the notion of events for enterprise modeling, knowledge representation and reasoning, information systems, and the semantic web, a widely applicable foundation for conceptual modeling requires a fuller account of the ontological notion of event. Figuring out different developing stages of event study would not only benefit its ontological research, but also pave the way for its application in interdisciplinary studies, such as event grammar and event calculus [4].

## C. Focusing on Events Grammar and Calculus

In this paper, the focus is on addressing the dynamic concepts, events, and negative events in systems. Specifically, we concentrate on events in what computer scientists would refer to as an *event grammar* and *event calculus*. Event grammar concerns the structure of possible event traces. An event trace represents an example of a particular execution of the system. Event traces can be effectively derived from the event grammar [10]. Event calculus represents an event that may occur in the world. The event calculus is utilized in the process that involves taking information about certain aspects of a scenario in the world and making inferences about other aspects of the scenario based on our knowledge of how the world works [11]. Event grammar and event calculus present well-developed topics that involve the notion of event. Analyzing the ways these areas of research approach the concept of event would further the understanding of events and their conceptions in systems and software engineering.

## D. Main Contribution in the Paper

Our contribution in this paper is to utilize a new definition of events. In computer science, there are many definitions of event, such as anything that happens, change in the properties of objects, and occurrence of and transition between states. This paper is based upon different conceptualizations using the thinging machine (TM) model and Lupascian logic to define *negative events*. An event is defined as a time-penetrated domain's region that is described in terms of things and five-action machines. Accordingly, samples from event grammar and event calculus are remodeled and analyzed in terms of this definition. The results point to an enriched modeling technique with an enhanced conceptualization of events that can benefit behavior modeling in systems.

## E. Paper Structure

This paper is structured as follows. The next section gives a brief description of the TM model with some new contributions. Section 3 includes a UML-based model that is remodeled to illustrate the TM approach and its basic notions. Section 4 gives a sample of current methods of representing events diagrammatically. The remaining sections of the paper involve remodeling examples from works in the areas of event grammar and event calculus.

## II. THE THINGING MACHINE MODEL

The TM model [12] is a semantic account of the ways expressions of the TM diagrammatic language relate to the things and actions comprising the world about. Entities in such a model are conceptualized as *thing/machine* (thimacs). Thimacs-ness is an arbitrary process that views entities as thimacs. A thimac represents a type of "state of affairs," similar to Minsky's frames [13]. For example, we might have a thimac representing a company with subthimacs called department, inventory, landlord, name, and address. A particular company is then to be represented by an instance of this thimac, obtained by inserting a time subthimac, as will be illustrated in the next section.

TM modeling has two levels of specification:
- A static model that represents static entities and static actions, using five specific static actions: create, process, release, transfer, and release
- A dynamic model that includes a static model subdiagram (called a *region*) and time, leading to the construction of events (i.e., the realization of static entities and actions)

Only thimacs that can embed time are realizable at the dynamic level. Thus, for example, a "square circle" is a static thimac that cannot be injected with time to be included in the dynamic model. Static thimacs have unique names; if there are two static thimacs with the same name, then they are injected with different time slots to be realizable at the dynamic level. Contradictory thimacs (regions) occupy the static level, described in philosophy as an "inconsistent multiplicity," but only consistent regions are realized at the dynamic level.

A thimac has a dual mode of being: the machine side and the thing side. The machine, called a TM, has the (potential) actions shown in Fig. 1. The sense of machinery originated in the TM actions indicating that everything that creates, processes, and moves (i.e., release, transfer, and receive) other things is a machine. Simultaneously, what a machine creates, processes (changes), releases, transfers, and/or receives is a thing. In this view, the "world" is a totality of thimacs.

Assemblages of thimacs may be formed from a juxtaposition of subthimacs that are bonded into a structure at a higher level at which they become parts. Thimacs comprise parts, which themselves are thimacs that comprise parts, and so on. Thimacs cannot be reduced to their parts because they (as wholes) have their own machines.

In general, we can view a thimac as a replacement of the notion of system, and the machine in Fig. 1 is a generalization of the famous input–process–output machine. A thimac is a whole that includes subthimacs that interrelate with each other and interact with the outside through the flow of things (thimacs).

TM modeling involves two levels that are characterized by staticity and dynamics. Staticity refers to a static model that represents the world of potentialities outside time where everything is present simultaneously. The dynamic model represents the world of actuality in time where events do not necessarily happen simultaneously. Time is a thimac (i.e., a thing that creates, processes, releases, transfers, and/or receives other things).

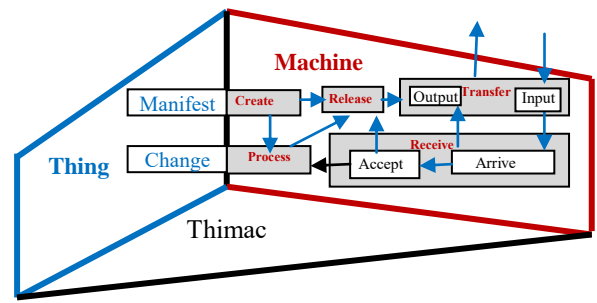

**Fig. 1. The thimac as a thing and machine.**

TM staticity and dynamics are applied not only to things but also to actions, thus we have static (potential) actions in the static model and dynamic (actual) actions in the dynamic TM model.

### A. The Machine

TM actions seen in Fig. 1 can be described as follows.

**Arrive**: A thing moves to a machine.

**Accept**: A thing enters the machine. For simplification, we assume that arriving things are accepted; therefore, we can combine **arrive** and **accept** stages into the **receive** stage.

**Release**: A thing is ready for transfer outside the machine.

**Process**: A thing is changed, handled, and examined, but no new thing results.

**Transfer**: A thing is input into or output from a machine.

**Create**: A new thing "becoming" (is found/manifested) is realized from the moment it arises (emergence) in a thimac. Simultaneously, it also refers to the "existence" of a thing, especially where we want to emphasize persistence in time as illustrated next.

Additionally, the TM model includes a *triggering* mechanism (denoted by a dashed arrow in this article's figures), which initiates a (nonsequential) flow from one machine to another. Multiple machines can interact with each other through the movement of things or through triggering. Triggering is a transformation from the movement of one thing to the movement of a different thing. The TM "space" is a structure of thimacs that forms regions of events (to be defined later). Note that for simplicity's sake, we may omit *create* in some diagrams, assuming the box representing the thimac implies its existence (in the TM model).

### B. Difference between Staticity and Dynamism

In TM modeling, staticity and dynamism are different from similarly named notions that are currently used in the literature. We briefly illustrate this difference by showing the TM modeling that corresponds to the distinction, in the *Stanford Encyclopedia of Philosophy* (summer 2020 edition) [14], between dynamic events, such as *John is walking*, and static events, such as *John is resting under a tree*.

Fig. 2 shows the static TM representation of the given sentences in one diagram. In Fig. 2, a person (pink number 1) creates and process of walking (2) and moves to (3) a spot under a tree (4) where he creates and the process of resting (5). The dynamic TM representation is built upon the notion of event. A TM event is constructed from a region (subdiagram of the static description) plus time. For example, Fig. 3 shows the event *John moves under the tree.* For simplification sake, events will be represented by their regions. Accordingly, Fig. 4 shows the dynamic representation that expresses the semantics of *John is walking*, and *John is resting under a tree*. The chronology of possible events is specified in the behavior model as shown in Fig. 5.

Note that the TM approach takes the side of philosophers (e.g., Whitehead) who conceive of objects as things that extend across time. Objects and events are things of the same kind [14]. The TM approach differs from such an ontological approach by introducing the notion of a thimac as a thing and a machine extended in time as illustrated in Fig. 5.

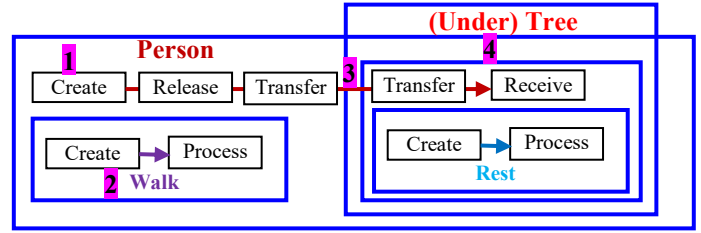

**Fig. 2** Static TM model.

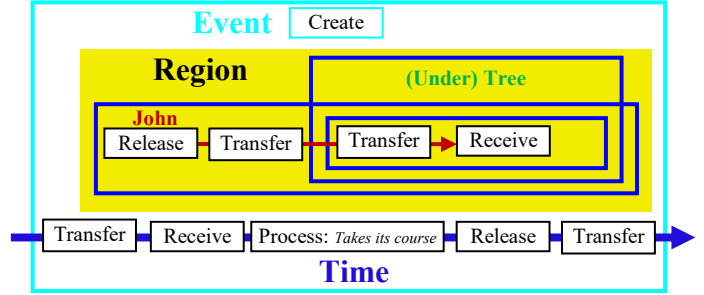

**Fig. 3** The event *John moves under the tree.*

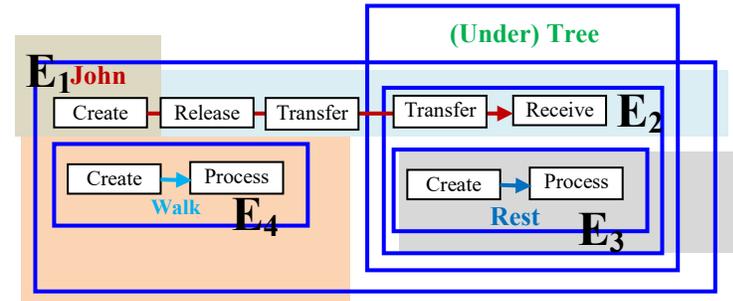

**Fig. 4** The dynamic TM model.

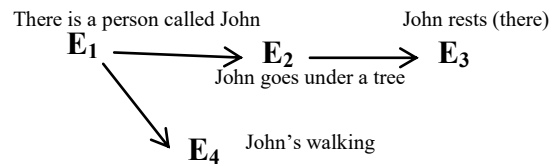

**Fig. 5** The TM behavior model.

### C. Two-Level Representation and Lupascian Logic

TM modeling is based on distinguishing two worlds: potentialities (also called *staticity* to avoid any human-related notions) described in terms of static actions (does not embed time) and actualities (dynamics), represented by (temporal) events. This implies that the activation of an event and its negative event can be defined as an alteration between the static level and the dynamic level. Instead of "process" versus "stop process," we have event (which moves the region to the dynamic level) versus "revert to static process," which returns the process to the static level. We take this method of eliminating negativity from the philosopher Stéphane Lupasco.

According to Brenner [15], every element *e* (in TM, an event, i.e., a thimac that contains a region plus time) is always associated with a *non-e* (in TM, a static thimac), such that the actualization of one entails the potentialization of the other and vice versa, alternatively, "without either ever disappearing completely" [15]. Lupasco provided a theoretical basis for the quasi-universal rejection of contradiction where every event is always associated with an anti-event, such that the actualization of an event entails the potentialization of a non-event and vice versa. This theoretical base is provided to emerge from a level of contradiction to the level of time attunement. Further illustration of this topic can be found in Al-Fedaghi [16]. The last section in this paper includes an example of negative events that are represented according to Lupascian logic.

A negative event is a return to potentiality from being an event. After the event *telephone call* ends with the negative event *there is no telephone call*); the caller, the person who receives the call, the telephone, etc., are still there, but they have returned to the non-eventness state. There are several types of negative events, and the ways to represent them in TM modeling is a topic for future research.

Note that entities (e.g., John) are viewed as a special type of event with TM expanded thimacs, which implies TM staticity persists through time. Regions with potentialities survive change in the dynamic level. Objects are always split and divided between a virtual domain of potencies and actuality manifested at any particular point in time [17]. Such a view does not imply the region (e.g., object) *already* contains what it *will* become (undermines the possibility of novelty [17]); rather, the static level includes pure capacity of every possibility. To preserve change at the dynamic level, actions (capacities) need to be actualized or they need to persist through time. Persistence of the thing called *John* implies that his *create*, *process*, *release*, *transfer*, and *receive* stages persist, as illustrated in Fig. 6. At the actual level, not only physical thimacs (e.g., John) occupy the four-dimension world, but also the three-dimensional world of creating, processing, releasing, transferring, and receiving. Such a topic deserves more research; here, the aim is to complete the TM picture about eventizing at the dynamic level.

## III. UML VS. TML MODELING

This section includes a UML-based model that will be remodeled using the TM approach.

According to Calafat [18], the UML conceptual model must include all relevant static and dynamic aspects of its domain. The terms *static* and *dynamic* are "defined" by showing diagrams from both types. The part of a conceptual description that deals with static aspects is called the structural schema, and the part that deals with dynamic aspects is called the behavioral schema. The structural schema defines the state of the domain that must be represented for the system to perform the memory function. A structural schema consists of taxonomy of entity types together with their attributes and taxonomy of relationship types among entity types [18]. Calafat [18] gives the UML conceptual schema that is shown in Fig. 7 as well as the behavioral schema corresponding to the structural schema as shown in Fig. 8.

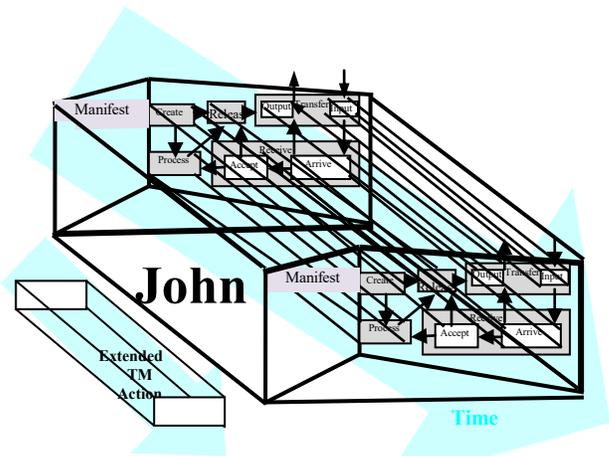

**Fig. 6. Illustration of *John* as an extended thimac in time.**

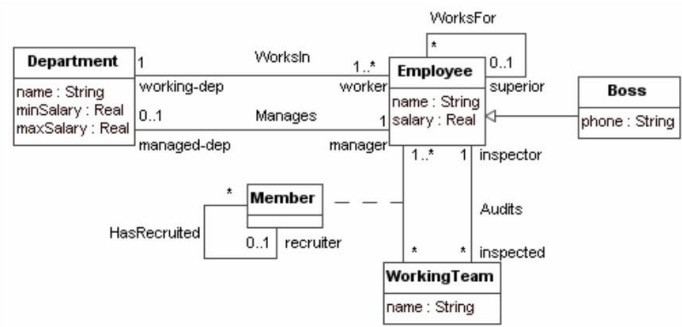

**Fig. 7 A structural schema for the domain of employees and their assignment to departments (from Calafat [18]).**

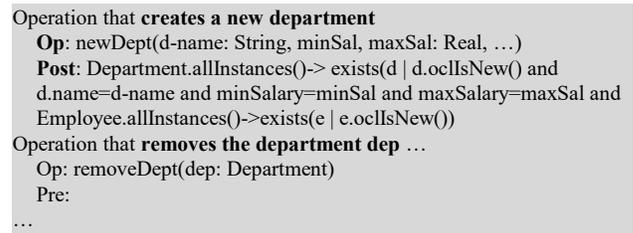

**Fig. 8 The behavioural schema corresponding to the structural schema (from Calafat [18]).**

Such a specification of an information system is suitable to illustrate the TM modeling and contrast it with a UML-based conceptual modeling. Of course, for space consideration, we need to develop only fragments of the system that partially demonstrates the difference between the two modeling methodologies, UML and TM.

### A. Static TM Model

As shown in the TM static model of Fig. 9, we select the following for inclusion in the model:

- Only *department*, *employee*, and *team*
- Relationships among classes: *work-in* and *manage*
- *Member* (of team)
- *Inspector* (of team).

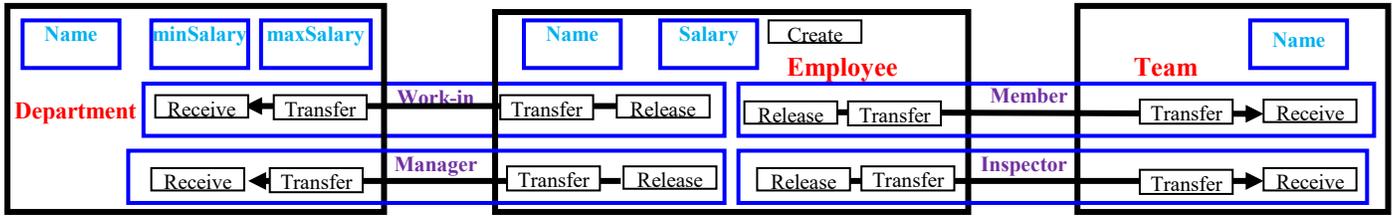

**Fig. 9 Partial static TM model.**

Fig. 9 gives only a general picture of the static TM model. More details of TM static modeling will be given later when we look more closely at portions of the model. Note that the static description in Fig. 9 includes actions (e.g., create and transfer) as potentialities that may be actualized at the dynamic level. We also deleted *create* operations in some thimacs under the assumption that the rectangle indicates their existence in the model.

### B. Dynamic TM Model

To demonstrate the dynamic TM modeling, we need to identify the events that are superimposed over the static model. Fig. 10 shows the definition of the event *Creating a new department*. For simplification, events will be represented by their regions. Fig. 11 shows regions of the sample events:

*Creating a new department*
*Creating a new employee*
*Assigning an employee to a department*
*Assigning an employee to a team*

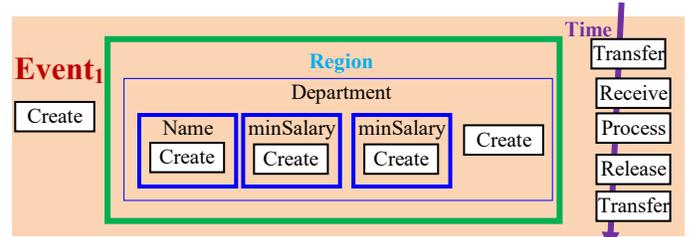

**Fig. 10 The event *Creating a new department*.**

### C. Granularity of the TM Model

Further granularity of the TM model is achieved using the same notions based on the TM. To save space, we will give only the dynamic TM models because it is possible to extract static TM models from dynamic TM models. Hence, we will discuss the following events in detail: the event (**Event₁** in Fig. 11) *Creating a new department* and the event *Deleting a department* (assuming that the department is empty, i.e., has no employees).

Fig. 12 shows further details of **Event₁** in terms of six more elementary events.

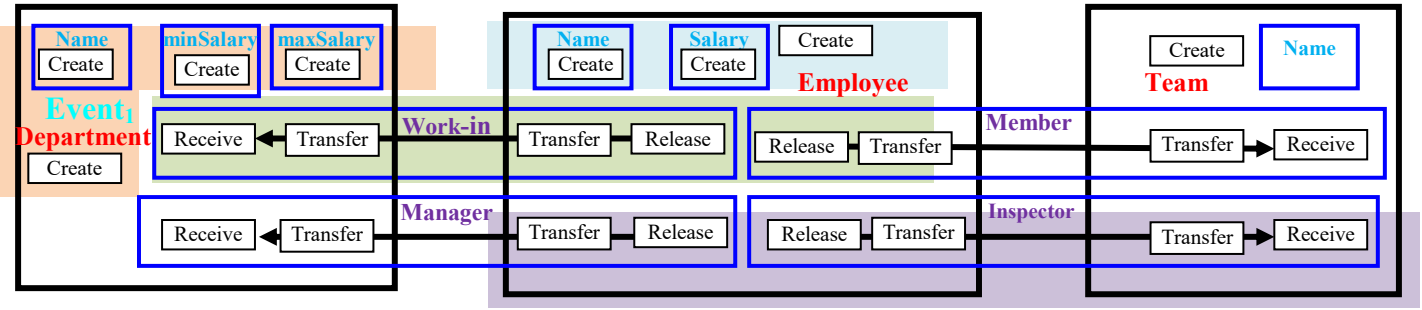

**Fig.11 Partial view of the dynamic model.**

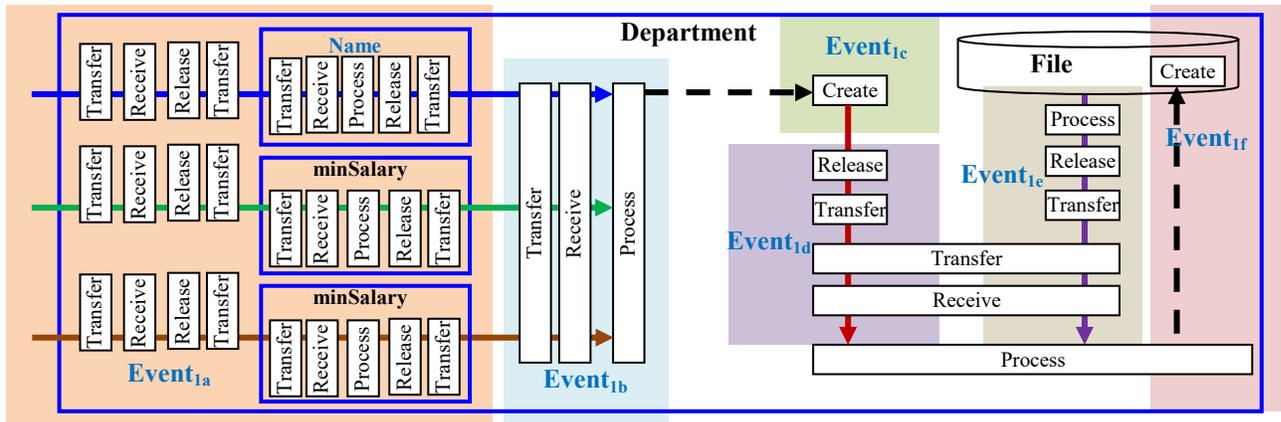

**Fig. 12 Details of the dynamic model of Event1, *Creating a new department*, in terms of more elementary events.**

We assume that *departments file* includes records of all departments.

Event$_{1a}$: Receiving the values of *Name* of department, *minSalary* and *MaxSalary* as external inputs

Event$_{1b}$: Processing the values of *Name*, *minSalary* and *MaxSalar*

Event$_{1c}$: Creating a new department record

Event$_{1d}$: Opening the *departments file*

Event$_{1e}$: Processing (a) the new record produced in Event$_{1c}$ and (b) the *departments file* to insert the record in the file

Event$_{1f}$: Creating a new *departments file* to replace the old one

Fig. 13 shows the behavioral model of *Creating a new department* in terms of the chronology of events.

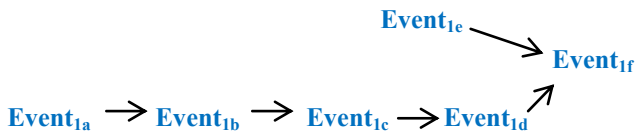

**Fig. 13 The behaviour model of *Creating a new department*.**

Fig. 14 shows further details of the *Deleting a department* behavioral model (not previously shown in Fig. 11), denoted as Event$_2$ in terms the following elementary events.

Event$_{2a}$: Receive the *name* of the department to be deleted

Event$_{2b}$: Extract a record from the *departments file*. Note that processing of the old *departments file* triggers the appearance (transfer-receive) of a record (i.e., does not create a record) because the record already exists inside the file.

Event$_{2c}$: Extract the name of the department in the extracted department record

Event$_{2d}$: Compare the two department names, (a) the one from input and (b) the one extracted from the department record

Event$_{2e}$: The two names are the same, so another department record is extracted (thus, skipping the addition of the record to the new *departments file*)

Event$_{2f}$: The two names are not the same, so the extracted record is added to the new version of the *departments file*

Fig. 15 shows the behavioral model of *Deleting a department* in terms of the chronology of events.

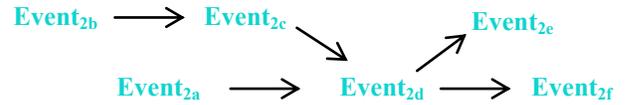

**Fig. 15 The behaviour model of *Deleting a department*.**

# IV. SAMPLE OF DIAGRAMMATIC EVENT REPRESENTATIONS

The remaining portion of this paper will cover events in the so-called *events grammar* and *events calculus*. However, in this section, we first give a current sample that illustrates the level of progress in representing events diagrammatically.

According to Wen et al. [19], grounding events into a precise timeline is important for understanding natural language, but it has received limited attention in recent work. They proposed a graph attention network-based approach to propagate temporal information over document-level event graphs constructed by shared entity arguments and temporal relations. The temporal representation allows a unified treatment of various types of temporal information and thus makes it convenient to propagate over multiple events. Wen et al. [19] give an example of an event graph (see Fig. 16) that represents the following description.

The <u>enemy</u> has now been **flown out** and we're treating them including a <u>man</u> who is almost dead with **a gunshot wound** to the chest after <u>we</u> (Royal Marines) **sent** in one of our <u>companies</u> of about 100 men in <u>here</u> (Umm Kiou) this morning. [19]

Entities in the text are underlined and events in the text are in boldface. Fig. 17 shows the corresponding static TM model.

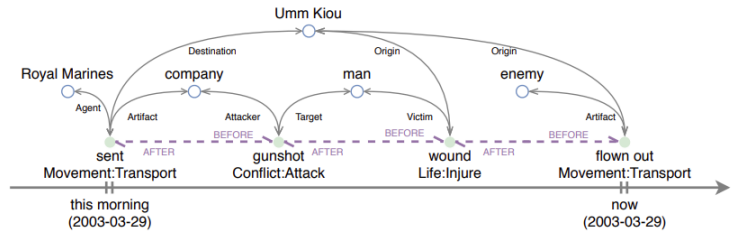

**Fig. 16 Solid lines are constructed from event arguments. Dashed lines are constructed from temporal relations (from Wen et al. [19]).**

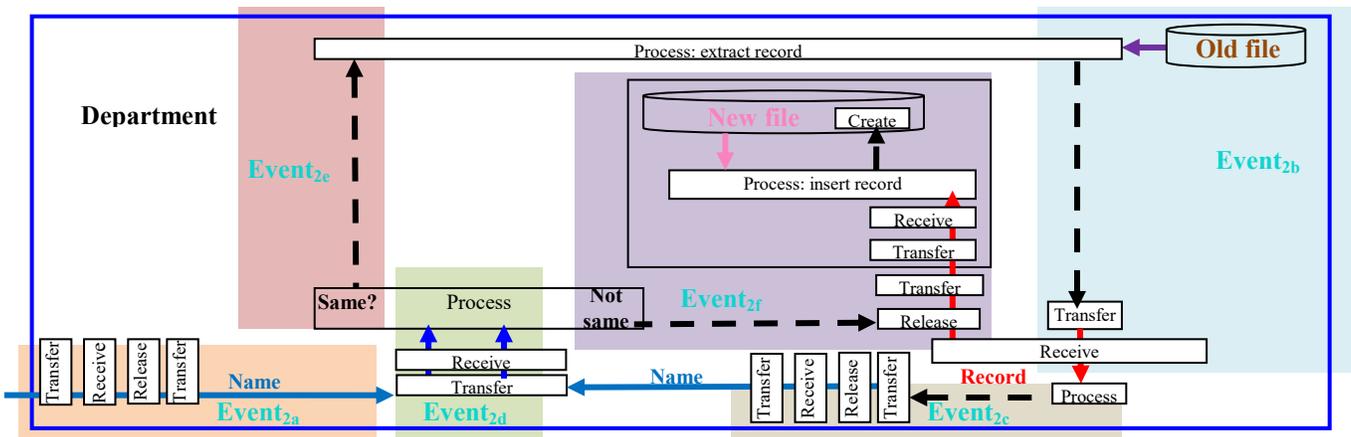

**Fig. 14 Details of deleting a department record**

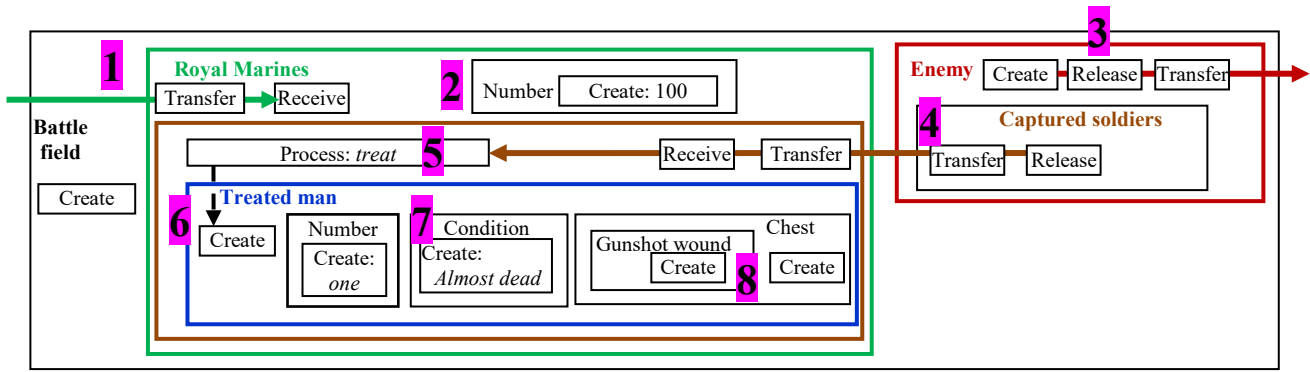

**Fig. 17 The static TM model.**

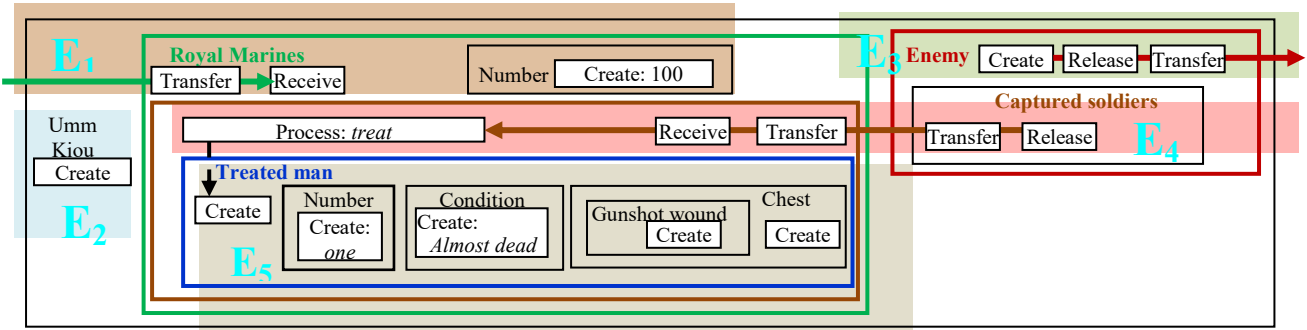

**Fig. 18 The dynamic TM model.**

First, the marines go to the battle field (pink number 1). Their number is 100 (2). The enemy flows out (3). The enemy's captured soldiers (4) are treated (5) by the marines. Out of the captured soldiers, a man is treated (6). The man is almost dead (7) with a gunshot wound to the chest (8).

We select the following list of events that are shown in the dynamic TM model of Fig. 18.

$E_1$: This morning, 100 royal marines were sent to battle field.
$E_2$: The battle field is Umm Kiou.
$E_3$: The enemy has been flown out.
$E_4$: The marines treated the captured soldiers.
$E_5$: A soldier is treated for a gunshot wound in the chest.

Fig. 19 shows the TM behavior model. The events that began with *this morning* is $E_3$ and *now* are the events $E_3$ and $E_4$. Wen et al.'s [19] diagram in Fig. 16 seems to be richer than the TM models are (e.g., destination, origin, and victim). The TM modeling is precise in terms of clear separation of events from the static description. We put the two representations side-by-side to be contrasted with respect to others features.

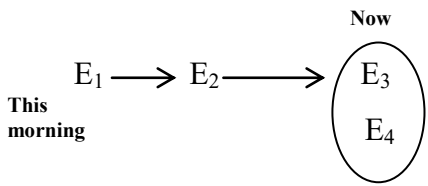

**Fig. 19 Behaviour TM model.**

## V. EVENT GRAMMAR

Augustan and Whitcomb [6] discussed ways to integrate the behavior of environment with behavior of system. They provided an example of an ATM_withdrawal schema that specifies a set of possible scenarios of interactions between the customer, ATM_system, and Data_Base. Each "event trace" generated from this schema can be considered a use case example [6]. See Figs. 20 and 21.

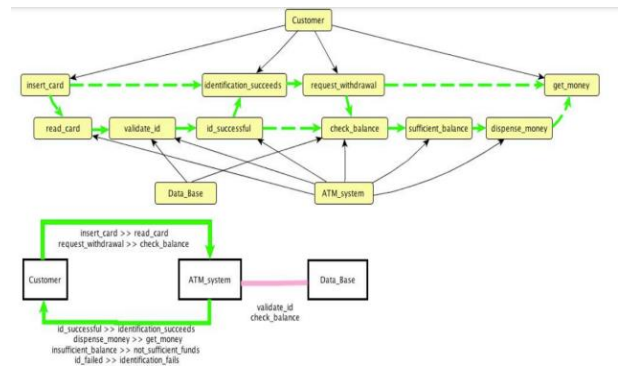

**Fig. 20 (Upper) an example of event trace for the ATM_withdrawal schema. (Lower) an architecture view for the ATM_withdrawal schema (from [6]).**

*Example  Withdraw money from ATM.*

```
SCHEMA ATM_withdrawal
ROOT Customer:        (*  insert_card
                          ( (  identification_succeeds
                               request_withdrawal
                          ( get_money | not_sufficient_funds ) ) |
                             identification_fails )                 *);
ROOT ATM_system:      (*  read_card        validate_id
                          ( id_successful  check_balance
                          ( sufficient_balance    dispense_money |
                            unsufficient_balance ) |
                          id_failed )                               *);
ROOT Data_Base:       (* ( validate_id | check_balance ) *);

Data_Base, ATM_system SHARE ALL validate_id, check_balance ;

COORDINATE    (* $x: insert_card *)           FROM Customer,
COORDINATE    (* $x: read_card *)             FROM ATM_system        ADD $x PRECEDES $y ;
COORDINATE    (* $x: request_withdrawal *)    FROM Customer,
              (* $y: check_balance *)         FROM ATM_system        ADD $x PRECEDES $y ;
COORDINATE    (* $x: identification_succeeds *) FROM Customer,
              (* $y: id_successful *)         FROM ATM_system        ADD $x PRECEDES $x ;
COORDINATE    (* $y: get_money *) FROM Customer,
COORDINATE    (* $x: dispense_money *) FROM ATM_system               ADD $x PRECEDES $x ;
COORDINATE    (* $x: not_sufficient_funds *)  FROM Customer,
              (* $y: unsufficient_balance *) FROM ATM_system         ADD $x PRECEDES $x ;
COORDINATE    (* $x: identification_fails *)  FROM Customer,
              (* $y: id_failed *)            FROM ATM_system         ADD $x PRECEDES $x ;
```

**Fig. 21 Specification of withdraw money from ATM (from [6]).**

Fig. 22 shows the static TM modeling of the withdrawal of money from an ATM. First, the customer inserts his or her card (pink number 1) that goes to the ATM (2), where it is processed (3) to trigger (4) the extraction of the customer's ID (5). The ID is sent to the bank (6) where it is processed (7). Assuming a valid ID, an approval note is sent to the ATM (8). The ATM then sends a request to the customer about the type of operation he or she wants to execute (9). We assume that the customer selects the withdrawal operation (10). The selection triggers (11) the ATM to send a request for the desired amount of money (12).

The customer inputs the amount (13) that is received (14) by the ATM to trigger sending the amount (15) to the bank to check whether sufficient funds exist in the customer's account (16). Assuming the fund is sufficient, a confirmation message is sent to the ATM (17). Upon receiving the message (18), the ATM triggers the cash management system to release (19) cash to the customer.

Accordingly, we designed the dynamic model to contain the following events as shown in Fig. 23.

$E_1$: A debit card is inserted for the ATM to extract the customer's ID.

$E_2$: The ID is sent to the database.

$E_3$: The ID is approved.

$E_4$: The ATM requests that the customer specify the required operation.

$E_5$: The customer specifies the withdrawal operation.

$E_6$: The ATM requests that the customer specify the amount.

$E_7$: The customer inputs the amount.

$E_8$: The ATM sends the amount to the bank.

$E_9$: The bank replies that the customer has sufficient funds in their account.

$E_{10}$: The ATM releases cash to the customer.

Fig. 24 shows the behavior model of the withdrawal operation. The red, blue, and black events indicate events that are performed by the customer, ATM, and bank, respectively. Again, it is difficult to compare Auguston and Whitcomb's [6] model and the TM model feature-by-feature. However, at this stage of study, it is sufficient to contrast them side-by-side. Future versions of each methodology may benefit from features in the other approach.

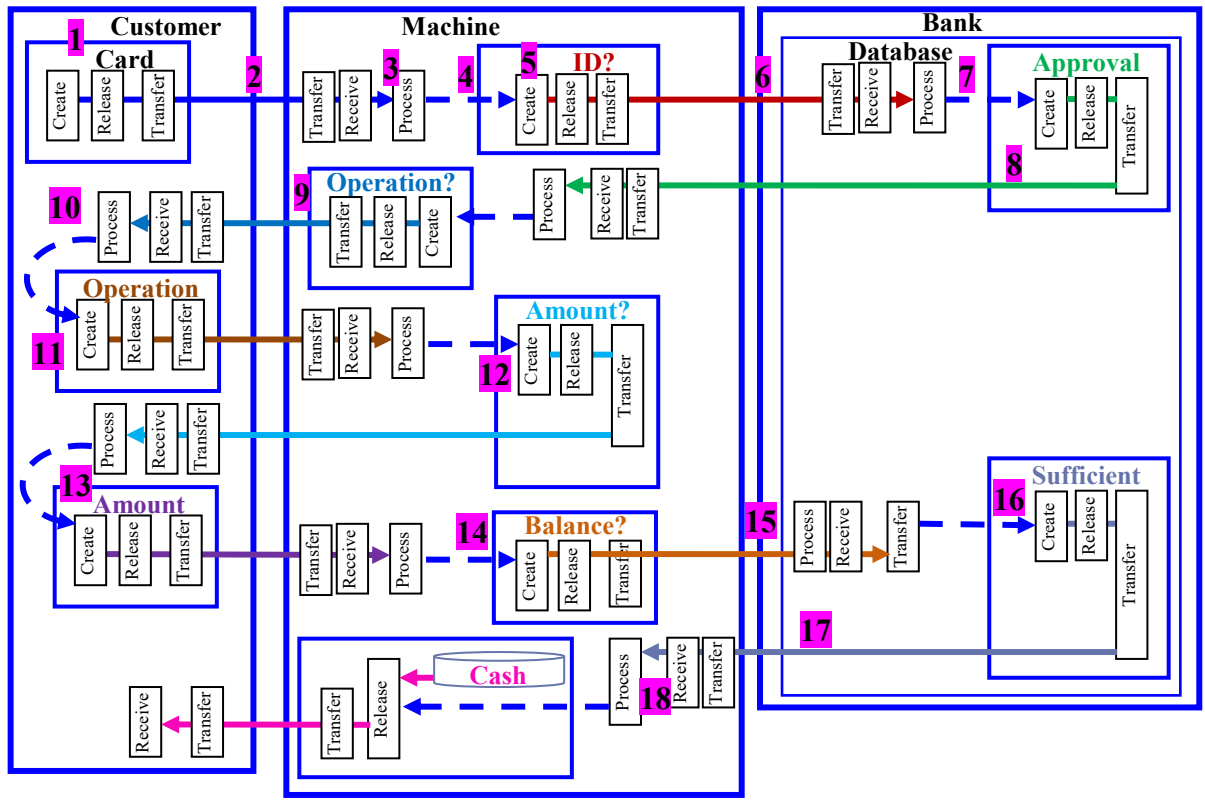

**Fig. 22 The static TM model of the ATM withdrawal operation.**

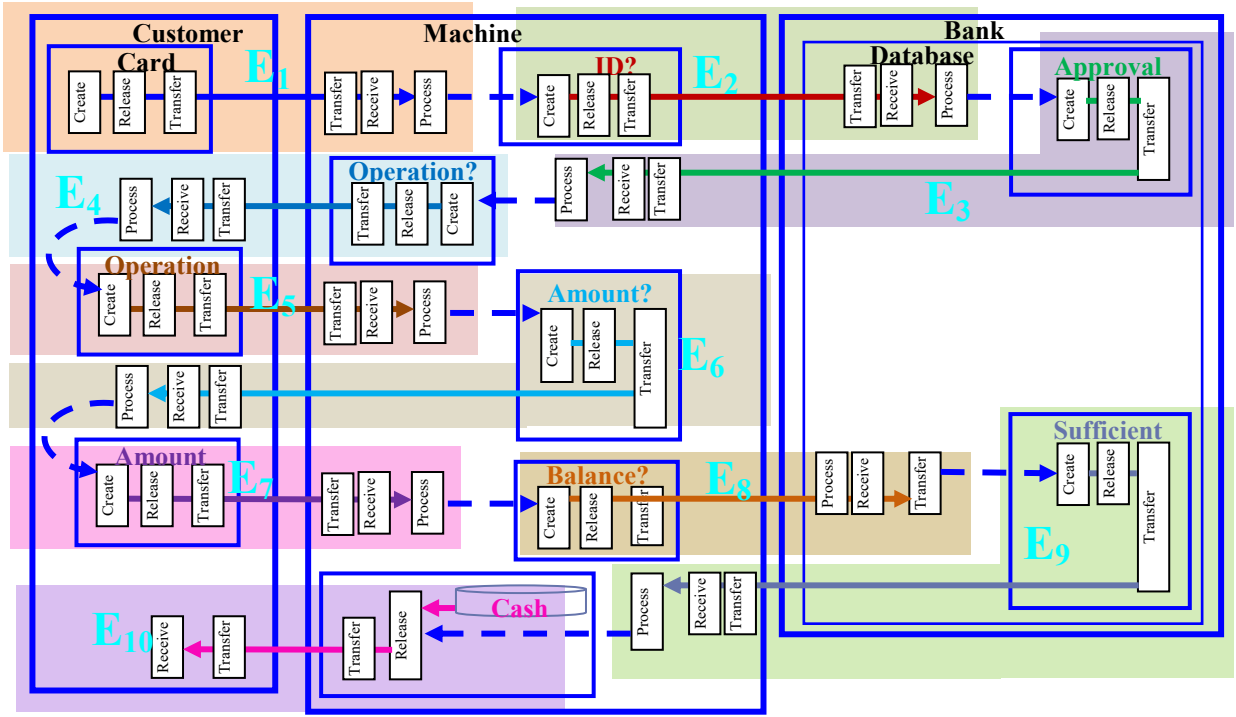

**Fig. 23 The static TM model of the ATM withdrawal operation.**

E₁ ➤ E₂ ➤ E₃ ➤ E₄ ➤ E₅ ➤ E₆ ➤ E₇ ➤ E₈ ➤ E₉ ➤ E₁₀

**Fig. 24 The behaviour model of the withdrawal operation.**

After this display of the ATM model in both Auguston and Whitcomb's [6] events grammar and the TM approach, it seems that the TM definition of a notion of an event is more ontologically complete than Auguston and Whitcomb's [6] notion of an event as "an abstraction of activity." An abstraction of activity begs the question of *what an activity is*. It is clear that such contrasting concepts of an event and its diagrammatic representation would benefit from development in both approaches. It is possible to adopt the idea of grammaticalizing events in TM from event grammar. Hence, the variation in the analyzed methodologies can lead to a better understanding of events. The next section presents one more problem that is studied in both event grammar and TM.

## VI. Event Grammar: Car Race

Whitcomb et al. [20] presented behavior modeling with a view of the architecture as a high-level description of possible systems behavior, emphasizing interactions between subsystems. Their event grammar provides a view of the behavior as a set of actions (event trace) with two basic relations, where the *precedes* relation captures the dependency abstraction and the *in* relation represents the hierarchical relationship. Whitcomb et al. [20] used event grammars as production grammars to generate instances of event traces, as in the following example of modeling a car race (see Fig. 25).

A sample specification of the problem of representing a car race is as follows.

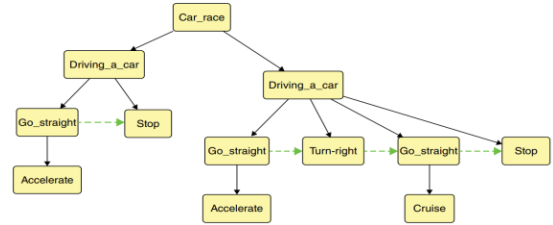

**Fig. 25 An event trace derived from the event grammar in the example (from Whitcomb et al. [20]).**

car_race: {+ driving_a_car +}; driving_a_car: go_straight (* ( go_straight | turn_left | turn_right ) *) stop; go_straight: ( accelerate | decelerate | cruise ); [20]

Fig. 26 shows the corresponding dynamic TM model. The static model can easily be extracted from Fig. 26.

We identify the following events:

E₁: The car is in the stop state.

E₂: The car transferred the stop state.

E₃: The car enters the forward state.

E₄: The car moves forward.

E₅: The car transfers from the forward state (pointing to different direction).

E₆: The car turns to the right.

E₇: The car turns to the left.

E₈: The car transferred from the turning left state.

E₉: The car transferred from the turning right state.

E₁₀: The car accelerates.

E₁₁: The car decelerates.

Note that these events are given in arbitrary sequence. The chronology of events is given in Fig. 27 which shows the TM behavior model of deriving a car in a car race.

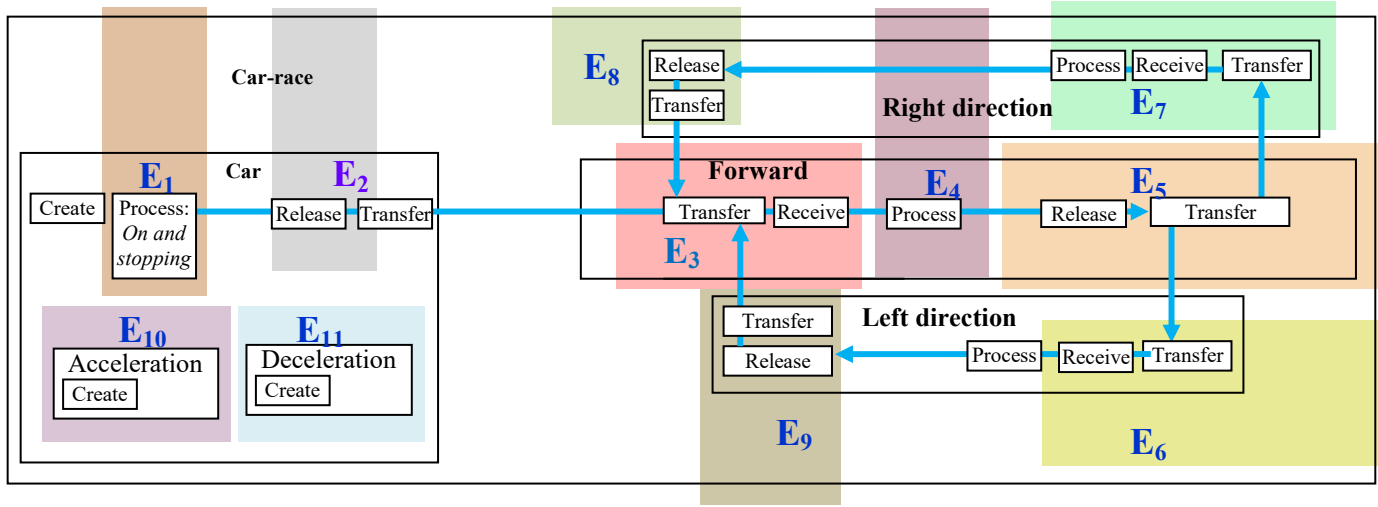

**Fig. 26 The dynamic TM model of a car race.**

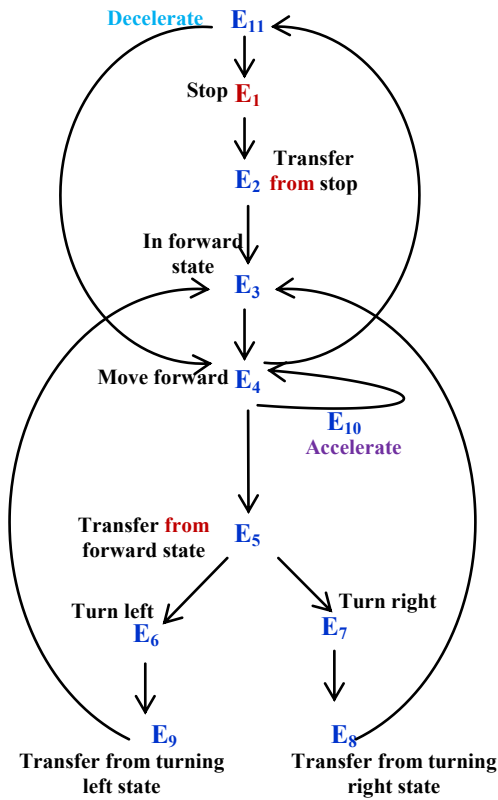

**Fig. 27 The behaviour TM model of driving a car in a race.**

Contrasting Fig. 27 with Whitcomb et al. [20]'s Fig. 25, it seems that the TM behavior model reflects a closer behavior image of chronology of events while Fig. 25 to present a hierarchical structure of events. We can conclude that showing the two methodologies side-by-side can benefit development of event modeling through a unified approach that adopts features from both of them.

## VII. EVENT CALCULUS

The event calculus domain utilizes logic language to represent events. In this section, the behavior of a phone system that uses event calculus will be presented.

According to Mueller [11], events are objects and statements can be made about the truth values of properties at time points and the occurrences of events at time points. The event calculus domain description consists of axiomatization, observations of world properties, and a narrative of known world events. Creating a domain description involves the following:

The logical constants for the particular problem are defined. These constants are used to specify observations of time-varying properties (fluents) and a narrative of event occurrences. Temporal ordering formulas, which relate the times of properties and events, may be specified. For example, it may be known that one event occurs prior to another event. Unique names axioms for events and fluents are specified [11].

According to Mueller [11], we perform complex commonsense reasoning about the effects of events whenever we use a telephone. The behavior of a phone is highly sensitive to context. If we pick up an idle phone, we expect to hear a dial tone, but if we pick up a phone that is ringing, we expect to be connected to the caller. We can represent our knowledge about telephones using positive and negative effect axioms. If an agent picks up an idle phone, the phone will have a dial tone and will no longer be idle:

$HoldsAt(Idle(p), t) \Rightarrow$ (3.1) $Initiates(PickUp(a, p), DialTone(p), t)$ $HoldsAt(Idle(p), t) \Rightarrow$ (3.2) $Terminates(PickUp(a, p), Idle(p), t)$

After picking up a phone, an agent may decide not to place a call. If an agent sets down a phone with a dial tone, the phone will be idle and will no longer have a dial tone:

HoldsAt(DialTone(p), t) ⇒ (3.3) Initiates(SetDown(a, p),Idle(p), t) HoldsAt(DialTone(p), t) ⇒ (3.4) Terminates(SetDown(a, p), DialTone(p), t)

Fig 29 shows the dynamic TM model of such a telephone system that is described as follow.

> If an agent picks up an idle phone, the phone will have a dial tone and will no longer be idle. After picking up a phone, an agent may decide not to place a call. If an agent sets down a phone with a dial tone, the phone will be idle and will no longer have a dial tone. When phone $p1$ has a dial tone and an agent dials phone $p2$ from $p1$, what happens depends on the state of $p2$. If $p2$ is idle, then $p1$ will be ringing $p2$, $p1$ will no longer have a dial tone, and $p2$ will no longer be idle. [11]

The new aspect in Fig. 28 is the representation of negative events (e.g., *stops a dial tone* and *not idle*). As mentioned in Section 2, the TM approach adopts the method of eliminating negativity from the philosopher Stéphane Lupasco. The negativity of a TM event ($E_i$) means returning to the static mode of representation; that is, being a region (denoted as $R_i$) without time element (see Al-Fedaghi [16]).

In Fig. 29, a negative event that corresponds to $E_i$ will be denoted as $R_i$ (R stands for region) is represented by a diamond-head arrow ( �does ➤◆ ). Accordingly, we have the following events:

$E_1$: A user picks up an idle telephone.
$R_1$: The phone is no longer idle.

$E_2$: The phone has a dial tone.
$E_3$: The telephone is set down with a dial tone.
$R_2$: The phone has no dial tone.
$E_4$: The phone is idle.
$E_5$: A number is dialed.
$E_6$: A signal is sent.
$E_7$: A signal is received.
$E_8$: The telephone rings (waiting to be picked up on the other end).
Fig. 29 shows the behavioral model of the telephone system.

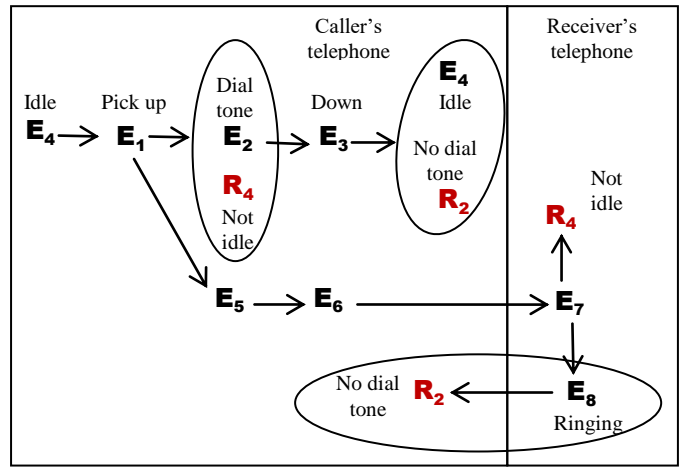

**Fig. 29 The TM behavior model.**

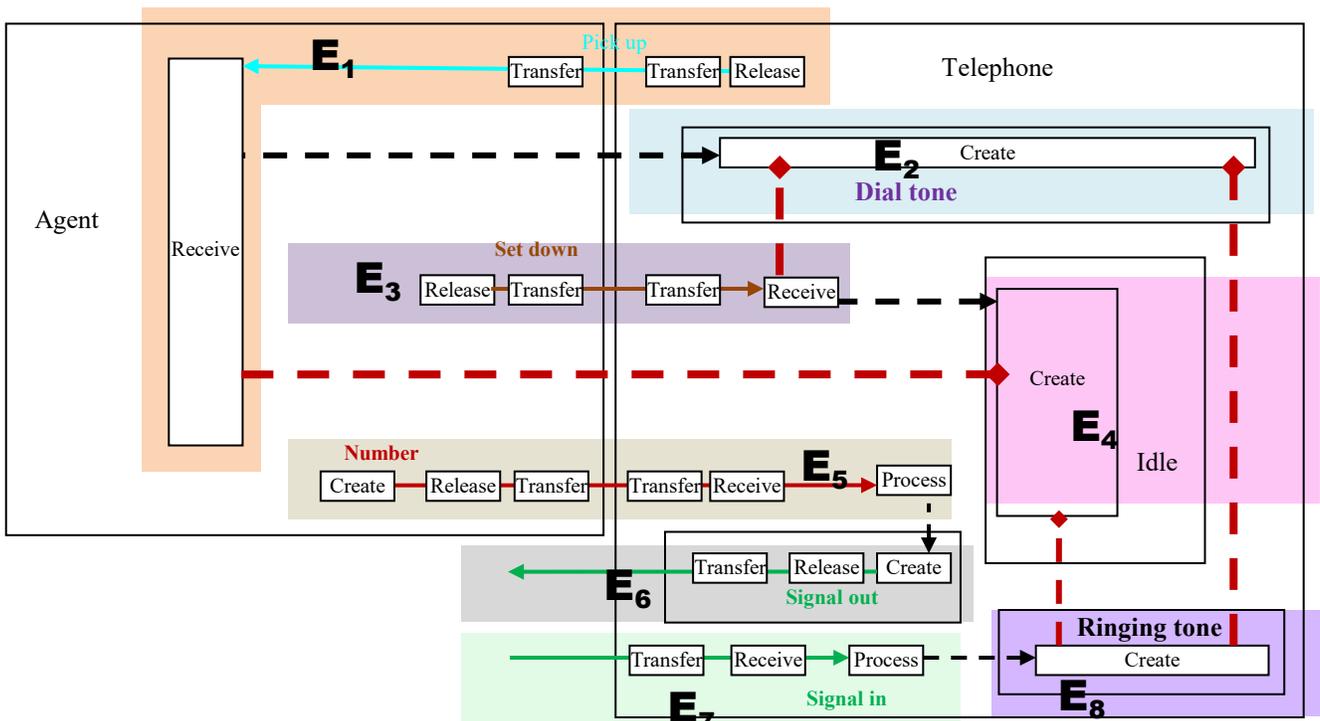

**Fig. 28 The dynamic TM model.**

Additional, we can use the TM representation of negative events for the absence of an event or refraining from doing action (e.g., the negative event *John does not pick up the telephone*). This would preserve *agency* in negative events. Studying TM representation of different types of negative events is a topic for further research.

## VIII. CONCLUSION

In this paper, we have studied the ontological foundations of conceptual modeling by addressing the concept of event. The study is based on TM using Lupascian logic to define negative events. A TM *event* is viewed as a time-penetrating region in the domain that is described in terms of things and five-action machines. Using such a definition, samples from an event grammar or event calculus have been remodeled and have been analyzed using the TM approach.

In conclusion, it seems that the TM definition of the notion of an event is a thought-provoking turn in comparison to the ones adopted in event grammar or event calculus. Additionally, it seems that the TM model has a broader foundation underpinned with such concepts as two-level representation and negative events. Although the research is on-going process, the results suggest that the TM diagrammatic modeling exposes inspiring multidisciplinary studies of notions such as negative events that spread to many research topics.

One purpose of incorporating TM modeling in different types of problems is to assess its viability. Hence, continuing to apply and introduce it to different problems in various subjects is a way to judge its value, if any, to modeling.